\documentstyle[12pt]{article}
\begin{document}

\begin{center}

{\bf Jet calculus problems of the perturbative quantum chromodynamics}\\

\vspace{2mm}

I.M. Dremin\footnote{Email: dremin@lpi.ru}\\

Lebedev Physical Institute, Moscow\\

\end{center}

\begin{abstract}

The perturbative quantum chromodynamics (pQCD) has been extremely successful
in the prediction and description of main properties of quark and gluon jets.
There are, however, some problems of the jet calculus with the higher order
corrections of the modified perturbative expansion which should be resolved
to get more precise statements. Some of them are discussed here.
\end{abstract}

The numerous achievements of pQCD in the jet calculus are well known and
described in the book \cite{5} and many review papers (see, e.g.,
\cite{dre1, koch, dgar, kowo, dr02}).
The leading approximation is perfect and only high order terms need more care.
Here I present a critical survey of some problems related to these calculations and
rarely discussed. The Figures demonstrating the comparison with experiment are
omitted to shorten the presentation. They can be found in the abovecited
review papers.

I mention briefly the following five problems:

1. Different characteristics of jets are differently sensitive to higher
order corrections. Therefore, for the comparison with experiment, one should
choose those which are not overshadowed by the leading terms of the perturbative
expansion and help most efficiently elucidate these corrections.

2. The correction terms are proportional to higher powers of the coupling
strength but can get the large numerical coefficients in front of them.
Thus, even though in asymptotics this expansion is valid due to the running
nature of the coupling strength, at present energies it could fail to provide
small corrections. Therefore one should find such characteristics where these
coefficients are small enough for corrections to be trusted.

3. It is very desirable to get the physical interpretation and motivation for
the value and nature of the higher order corrections (especially, for cumulant
moments).

4. The QCD equations or approximations used in the jet calculus are sometimes
not completely precise themselves. Their modifications can be considered or 
the influence of the omitted terms estimated.

5. Some shortcomings of the analytic approach and numerical solutions are
discussed.

I will be mainly concerned with jet characteristics in some sense related to
the jet multiplicity distributions that are closer to my personal interests.
First, let me remind some simplest definitions \cite{5, dre1} concerning jet
multiplicities in QCD. The generating function $G$ is defined by the formula
\begin{equation}
G(y,u) = \sum_{n=0}^{\infty }P_{n}(y)u^{n}  ,                     \label{3}
\end{equation}
where $P_{n}(y)$ is the multiplicity distribution at the scale
$y=\ln (p\Theta /Q_0 )=\ln (2Q/Q_{0})$, $p$ is the initial momentum, $\Theta $
is the angle of the divergence of the jet (jet opening angle), assumed here to
be fixed, $Q$ is the jet virtuality,  $Q_{0}=$ const, $u$ is an auxiliary
variable which is often omitted to shorten notations. The analytic properties
of the generating functions in $u$ are of the special interest (see
\cite{dre1, dgar}) in view of
some analogies with the statistical physics, but we will not consider them here.

The moments of the distribution are defined as
\begin{equation}
F_{q} = \frac {\sum_{n} P_{n}n(n-1)...(n-q+1)}{(\sum_{n} P_{n}n)^{q}} =
\frac {1}{\langle n \rangle ^{q}}\cdot \frac {d^{q}G(y,u)}{du^{q}}\vline _{u=1}, 
\label{4}
\end{equation}
\begin{equation}
K_{q} = \frac {1}{\langle n \rangle ^{q}}\cdot \frac {d^{q}\ln G(y,u)}{du^{q}}
\vline _{u=1}. \label{5}
\end{equation}
Here, $F_q$ are the factorial moments, and $K_q$ are the cumulant moments, 
responsible for total and genuine (irreducible to lower ranks) correlations,
correspondingly. These moments are not independent. They are connected by
definite relations which can easily be derived
from moments definitions in terms of the generating function:
\begin{equation}
F_{q} = \sum _{m=0}^{q-1} C_{q-1}^{m} K_{q-m} F_{m} .              \label{11}
\end{equation}

The QCD equations for the generating functions are\footnote{To exclude the
nonperturbative region from further consideration, the limits of integration
in these equations are often chosen as exp$(-y)$ and 1- exp$(-y)$ which tend
to 0 and 1 at high energy $y$.}:
\begin{eqnarray}
&G_{G}^{\prime }&= \int_{0}^{1}dxK_{G}^{G}(x)\gamma _{0}^{2}[G_{G}(y+\ln x)G_{G}
(y+\ln (1-x)) - G_{G}(y)] \nonumber \\ 
&+&n_{f}\int _{0}^{1}dxK_{G}^{F}(x)\gamma _{0}^{2}
[G_{F}(y+\ln x)G_{F}(y+\ln (1-x)) - G_{G}(y)] ,   \label{50}
\end{eqnarray}
\begin{equation}
G_{F}^{\prime } = \int _{0}^{1}dxK_{F}^{G}(x)\gamma _{0}^{2}[G_{G}(y+\ln x)
G_{F}(y+\ln (1-x)) - G_{F}(y)] .                                   \label{51}
\end{equation}
Here $G^{\prime }(y)=dG/dy ,$ $ n_f$ is the number of active flavours,
\begin{equation}
\gamma _{0}^{2} =\frac {2N_{c}\alpha _S}{\pi } .               \label{52}
\end{equation}
The running coupling constant in the two-loop approximation is
\begin{equation}
\alpha _{S}(y)=\frac {2\pi }{\beta _{0}y}\left( 1-\frac {\beta _1}
{\beta _{0}^{2}}\cdot \frac {\ln 2y}{y}\right)+O(y^{-3}), \label{al}
\end{equation}
where
\begin{equation}
 \beta _{0}=\frac {11N_{c}-2n_f}{3}, \;\;\;\;\;\;
 \beta _1 =\frac {17N_c^2-n_f(5N_c+3C_F)}{3}.
 \label{be}
\end{equation}
The labels $G$ and $F$ correspond to gluons and quarks,
and the kernels of the equations are
\begin{equation}
K_{G}^{G}(x) = \frac {1}{x} - (1-x)[2-x(1-x)] ,    \label{53}
\end{equation}
\begin{equation}
K_{G}^{F}(x) = \frac {1}{4N_c}[x^{2}+(1-x)^{2}] ,  \label{54}
\end{equation}
\begin{equation}
K_{F}^{G}(x) = \frac {C_F}{N_c}\left[ \frac {1}{x}-1+\frac {x}{2}\right] ,   
\label{55}
\end{equation}
$N_c$=3 is the number of colours, and $C_{F}=(N_{c}^{2}-1)/2N_{c}
=4/3$ in QCD.  

Herefrom, one can get equations for any moment of the multiplicity distribution
both for quark and gluon jets. One should just equate the terms with the same
powers of $u$ in both sides of the equations.
In particular, the equations for average  multiplicities read
\begin{eqnarray}
\langle n_G(y)\rangle ^{'} =\int dx\gamma _{0}^{2}[K_{G}^{G}(x)
(\langle n_G(y+\ln x)\rangle +\langle n_G(y+\ln (1-x)\rangle -\langle n_G(y)
\rangle ) \nonumber  \\
+n_{f}K_{G}^{F}(x)(\langle n_F(y+\ln x)\rangle +\langle n_F(y+
\ln (1-x)\rangle -\langle n_G(y)\rangle )],  \label{ng}
\end{eqnarray}
\begin{equation}
\langle n_F(y)\rangle ^{'} =\int dx\gamma _{0}^{2}K_{F}^{G}(x)
(\langle n_G(y+\ln x)\rangle +\langle n_F(y+\ln (1-x)\rangle -\langle n_F(y)
\rangle ).   \label{nq}
\end{equation}

Their solutions can be looked for as
\begin{equation}
\langle n_{G,F}\rangle \propto \exp (\int ^{y}\gamma _{G,F}(y\prime )dy\prime ).
\label{57}
\end{equation}
The lower limit of integration has not been fixed because its variation
results in the substitution of a new normalization constant which is not
shown in the above relation but is in practice considered as a fitted parameter
which depends on the nonperturbative component of the underlying dynamics of a
process.

Using the perturbative expansion of the exponent in (\ref{57})
\begin{equation}
\gamma _G \equiv \gamma = \gamma _{0}(1-a_{1}\gamma _{0}-a_{2}\gamma _{0}^{2}-
a_3\gamma _0^3)+O(\gamma _{0}^{5})
 , \label{X}
\end{equation}
one arrives to the so-called modified perturbative expansion of QCD. This
means that the perturbative expansion has been used in the exponent of the
expression for a physical quantity, i.e., even the first term includes higher
power corrections of the ordinary perturbative formulas. Moreover, the
expansion parameter is the coupling strength itself and not its squared value
$\alpha _S$ as usually happens. The structure of the equations (\ref{50}),
(\ref{51}) dictates such series. It was first shown in \cite{13} that the
systematic expansion can be obtained by considering the Taylor series at low
$x$ in Eqs (\ref{50}), (\ref{51}). There it was used for higher order
calculations in gluodynamics. The ordinary
perturbative expansion for mean multiplicity, if boldly
attempted, would surely fail because the coupling strength decreases with
energy while multiplicities increase mainly due to the enlarged phase space
volume. The coefficients $a_i$ are calculable from the eqs (\ref{ng}),
(\ref{nq}).

Let us briefly mention that the equations (\ref{50}), (\ref{51}) can be exactly
solved \cite{21, dhwa} for fixed coupling strength, i.e., if $\gamma _0$ is set
constant. Then the mean multiplicities increase like a power of energy.
The comparison with experiment has been done in recent paper \cite{cgar}.

For the running coupling strength the multiplicities increase
\cite{5, dg, cdnt8} slower than power-like but stronger than logarithmically,
namely
\begin{equation}
\langle n_{G,F}\rangle=A_{G,F}y^{-a_{1}c^2 }\exp ( 2c\sqrt y+
\delta _{G,F}(y)),                                  \label{mul}
\end{equation}
where $c=(4N_c/\beta _0)^{1/2}$,
\begin{equation}
\delta _G(y)=\frac {c}{\sqrt y}[2a_2c^2+\frac {\beta _1}{\beta _{0}^{2}}
(\ln 2y+2)]+\frac {c^2}{y}[a_3c^2-\frac {a_1\beta _1}{\beta _{0}^{2}}
(\ln 2y +1)])+O(y^{-3/2}).
 \label{mean}
\end{equation}
The corresponding expression for $\delta _F(y)$ can be easily obtained from
the formulas for $\gamma _F$.
Usually, in place of $\gamma _F$ the ratio of average multiplicities in gluon
and quark jets 
\begin{equation}
r=\frac {\langle n_G\rangle }{\langle n_F\rangle }=\frac {A_G}{A_F}
\exp(\delta _G(y)-\delta _F(y)) \label{rat}
\end{equation}
is introduced, and its perturbative expansion
\begin{equation}
r = r_0 (1-r_{1}\gamma _{0}-r_{2}\gamma _{0}^{2}-r_3\gamma _0^3)+
O(\gamma _{0}^{4})
  \label{Y}
\end{equation}
is used. The analytic expressions and numerical values of the parameters
$a_i, r_i$ for all $i\leq 3$ have been calculated from the perturbative
solutions of the above equations. All of them (except $r_0=N_c/C_F=9/4$) are
at least twice less than 1 (the review is given in \cite{dgar}).

The relation between the anomalous dimensions of gluon and quark jets is
\begin{equation}
\gamma _F=\gamma -\frac {r'}{r},  
\end{equation}
where
\begin{equation}
r'\equiv dr/dy=Br_0r_1\gamma _0^3[1+\frac {2r_2}{r_1}\gamma _0+
(\frac {3r_3}{r_1}+B_1)\gamma _0^2+O(\gamma _0^3)]       \label{derr}
\end{equation}
with $B=\beta _0/8N_c; \; B_1=\beta _1/4N_c\beta _0.$

Thus
\begin{equation}
\gamma _F = \gamma _{0}[1-a_{1}\gamma _{0}-(a_{2}+Br_1)\gamma _{0}^{2}-(a_3+
2Br_2+Br_1^2)\gamma _0^3-(a_4+B(3r_3+3r_2r_1+B_1r_1+r_1^3)) \gamma _{0}^{4}].
\label{gf}
\end{equation}

At present, there exist three approaches to treating multiplicities. 
In analytic solutions of the equations the perturbative approach with
approximate energy conservation is used. The numerical solution allows to 
account accurately for energy conservation. However, the transverse momentum
is taken into account only by angular ordering of jets in both approaches.
Both energy and momentum are conserved in Monte Carlo QCD models which provide
best fits to experimental data at present energies. Their predictions, however,
differ at higher energies mainly due to hadronization models used. That is why 
further studies are needed.\\

1. {\bf Sensitivity to high order terms.}
The experimental data about the energy dependence of mean multiplicity in
$e^+e^-$-annihilation are well described in all approaches.
The two leading terms in expressions (\ref{mul}) completely determine it.
They are the same for quark and gluon jets. That is why gluodynamics can
be used for their estimate as was done in early years.
The higher order corrections given by $\delta _{G,F}$ are almost 
unnoticeable there. Thus mean multiplicities are not sensitive to these
corrections by themselves.

However, if one considers their ratio $r$, it happens
to be really sensitive. This is because in the ratio the two leading terms
corresponding to leading order (LO) and next-to-leading order (NLO) cancel
since they are the same for both quark and gluon jets. Therefore, only
higher order corrections determine the energy behaviour of the ratio $r$.
The first term $r_0=9/4$ is given by the relative strengths of gluon
and quark forces. The next term is proportional to $\gamma _0$ and will
be called NLO$_r$-correction in distinction to common NLO-terms. Actually,
NLO$_r$ corresponds to 2NLO-terms (like $a_2$) because of the cancellation
of the NLO (power-like in $y$) terms in the ratio of multiplicities (see Eq.
(\ref{gf})). In the same sense the "$r_3$"-term in $r$ corresponds to 4NLO
contribution in $\gamma $ even though it is proportional to $\gamma _0^3$ etc
(see \cite{cdnt8}). This leads to shift and misuse of the terminology for the
anomalous dimensions $\gamma $'s and for the ratio $r$.

Thus we have found the characteristic which is more sensitive to higher 
order corrections than mean multiplicities. The experimental data about the
ratio $r$ are described with much lower accuracy about 15-20$\%$ in such
analytic approach (see \cite{cgar}). Even though each subsequent
perturbative term in $r$ improves the agreement, no precise fit has been
achieved yet. 

However, one should mention here that the computer solution of the equations
\cite{lo2, lo1} provides the quantitative fit. This indicates
that the higher order uncalculated corrections are still comparatively large
for this ratio up to the highest presently available energies. Thus the
discrepancy with analytic results is of a purely technical origin.

Another very sensitive characteristic is the behaviour of the factorial
moments (\ref{4}) as functions of the size of the phase space bins in which
they are measured. Here, one has to deal with a part of the phase space and
the above equations are not applicable directly. One has to use the Feynman
diagram technique for the treatment of these small bins \cite{owos, ddre, bmpe}.
This complicates the matter. It was impossible to account for high order
corrections. Some NLO-terms have only been considered in \cite{ddre}.
From comparison with experiment (see, e.g., \cite{ddki, kitt, sark})
it is seen that the qualitative behaviour
is described but quantitatively the disagreement becomes stronger at smaller
bins. This poses the problem of the proper account of higher order corrections.
Possible flow of partons from small bins should be considered more precisely.
The newly developed technique of the so-called non-global logarithms \cite{sala}
can be helpful in this respect.\\

2. {\bf High order coefficients.}
Fortunately, the coefficients $a_i$ and $r_i$ happened to be small enough
(see the Table in \cite{dgar}) so that the subsequent terms in the expansions
of $\gamma $ and $r$
can be trusted even at the rather large values of the expansion parameter
$\gamma _0\approx 0.4 - 0.5$ at present energies. This is not always the case
for some other characteristics.
If the high order terms become larger than 1, the expansion can not be trusted.
Thus the next problem is to find such characteristics for which it does not
happen. Only these features can be reliably compared with experiment.

This criterium becomes crucial, e.g., for the slope $r'$ of the ratio $r$.
The cancellation of two leading terms in the ratio $r$ reveals itself 
in the proportionality of the scale (energy) derivative $r'$ to $\gamma _0^3$.
Therefore it can be calculated up to the terms $O(\gamma _0^5)$. The leading
term is very small (about 0.02 at the Z$^0$-resonance). Asymptotically, all
corrections vanish. However, at present energies of Z$^0$, they are so large
that calculations become unreliable. The second term in the brackets in
(\ref{derr}) is larger than 1 since $2r_2/r_1 \approx 4.9$ and $\gamma \approx
0.45 - 0.5$. Even the third term is approximately about 0.4. The problem of
convergence of the series at Z$^0$-energies and below becomes crucial.

Therefore, it is desirable to use at present energies such characteristics
which are sensitive to these corrections and do not possess large coefficients
in front of the expansion parameter. In particular, it has been shown in
\cite{cdnt8} that these coefficients are smaller in the ratio of derivatives
(slopes)
\begin{equation}
r^{(1)}=\frac {\langle n_G\rangle '}{\langle n_F\rangle '}. \label{ratd}
\end{equation}
This ratio should be slightly larger than $r$
\begin{equation}
r^{(1)}\approx r(1+Br_1\gamma _0^2)\approx r(1+0.07\gamma _0^2).\label{r1}
\end{equation}
The same is true for the ratio of curvatures (or second derivatives)
\begin{equation}
r^{(2)}=\frac {\langle n_G\rangle ''}{\langle n_F\rangle ''}. \label{radd}
\end{equation}
It is even closer to the asymptotics
\begin{equation}
r^{(2)}\approx r(1+2Br_1\gamma _0^2)\approx r(1+0.14\gamma _0^2).\label{r2}
\end{equation}
The QCD predictions for them
\begin{equation}
r<r^{(1)}<r^{(2)}<2.25
\end{equation}
have been confirmed in experiment. 

The present experimental accuracy does not allow, unfortunately, to 
measure these values more accurately. As one sees, in expressions for 
$r^{(1)}, r^{(2)}$ the coefficients in front of $\gamma _0^2$ are 
slightly decreased compared with $r$ but not in front of $\gamma _0$.
The last ones cancel in their ratios to $r$ so that the second order
terms are left. However, these ratios $r^{(1)}/r$ and $r^{(2)}/r$
have not yet been accurately measured. Further search for such 
characteristics is needed. \\

3). {\bf Interpretation.}
Another question I'd like to raise concerns physical interpretation of
the high order effects. First of all I mean the oscillations of cumulant
moments as functions of their rank in QCD. They have not yet been completely
clarified. They were predicted analytically \cite{13} and numerically \cite{41}
as the effect of the high order terms of the modified perturbative expansion.
Their detailed study was recently performed
in \cite{bfoc} by the numerical solution of QCD equations. First experimental
confirmation was found in \cite{dabg, sld}.

A peculiar feature of multiplicity distributions has been noticed in \cite{bfoc}.
The even order factorial moments $F_2, F_4, F_8, F_{16}$ become equal to 1 at
the energy about 20 GeV. This implies that the distribution has a
quasi-Poissonian shape. At lower energies it is sub-Poissonian, at higher ones -
super-Poissonian. The similar conclusion for gluon jets can be derived from
results of \cite{cgar}. It has been stated that no analytic explanation of it
is known. Actually, it is hard to proceed with analytic calculations to high
rank moments because the expansion parameter $q\gamma $ becomes large. However
one can answer the question about the energy where $F_2$ is equal to 1 in
NLO-approximation. This moment plays the main role for the distribution since
other moments are quite small. The equality $F_2=1$ implies $K_2=0$, and 
according to \cite{13} can be written as
\begin{equation}
1-4h_1\gamma = 0,   \label{f21}
\end{equation}
where $h_1=11/24, \gamma $ is the QCD anomalous dimension. The energy $E$ at 
which this is satisfied is given by
\begin{equation}
\ln \frac {M_Z}{E}=\frac {2\pi}{\beta_0}\left (\frac {1}{\alpha _Z}-\frac {1}
{\alpha _E}\right )   \label{Ef2}
\end{equation}
with $\alpha _E=\pi \gamma ^2/6=\pi /96h_1^2; \; \alpha _Z\approx 0.118; \;
\beta _0=9$ for $n_f=3$. Herefrom one easily estimates
\begin{equation}
E\approx 20 \; GeV.    \label{Ech}
\end{equation}
Thus the analytic estimations of the transition region coincide quite well with
computer calculations \cite{bfoc} and experiment \cite{cgar}. Its universality 
for other collisions would be interesting to check.

Usually exploited phenomenological
distributions of the probability theory do not possess any oscillations.
E.g., all cumulant moments of the Poisson distribution are identically zero.
One interprets this as the absence of genuine correlations irreducible to
the lower-rank correlations. For the negative binomial distribution with the
parameter $k$ one easily gets
\begin{equation}
H_q=\frac {K_q}{F_q}=kB(q,k)>0.
\end{equation}
Since $F_q$ are always positive according to their definition, this inequality
implies the positive values of $K_q$.

In the leading order approximation, the gluodynamics equation for the 
generating function 
\begin{equation}
[\ln G(y)]''=\gamma _0^2(G(y)-1)
\end{equation}
transforms in the relation
\begin{equation}
q^2K_q=F_q \;\;\;\; or \;\;\;\; H_q=\frac {1}{q^2}.   \label{hqq2}
\end{equation}
However already in the next-to-leading order $H_q$-moments become negative 
with a minimum at the rank $q_{min}\approx \frac {24}{11\gamma _0}+0.5\approx 5$
\cite{13}. This minimum is rather stable. Nevertheless this is a purely
preasymptotic feature. The minimum slowly moves to higher ranks with
energy increase and disappears in asymptotics as is required according to the
formula (\ref{hqq2}). At higher orders of the perturbative expansion, the
oscillations of higher rank cumulant moments
show up \cite{41, bfoc}. They have been confirmed in experiment.
Convergence to $1/q^2$-limit with energy increase has been noticed in \cite{bfoc}
for low-rank moments.
We are interested to get from experiment the data about the energy
behaviour of the ratios $H_q$ or, better, of the asymptotically normalized
ratios $T_q=q^2H_q$ which should tend to 1 in asymptotics independently of $q$.
It would ask for high precision data at different energies.

The cumulants remind the virial coefficients of statistical physics.
The changing character of the genuine correlations (described by the cumulants)
implies that attraction (clustering) is replaced
by repulsion (and vice versa) in particle systems with different number of
particles. The similar behavior of correlators is known, e.g., in the theory
of super-fluidity. In superconductivity, it is at the origin of Cooper pairs.
Has it any impact on the hydrodynamical theory of multiple
production? It would be exciting to find other examples of such a behaviour
in hadronic systems. From experimental side, it would, perhaps, reveal itself
in the irregular behavior of mean multiplicities of subjets.\\

4). {\bf Generalization.}
Finally, there exists the problem of possible generalization of the equations
for the generating functions. As such, the Eqs (\ref{50}), (\ref{51}) have only
been proved (see, e.g., \cite{5}) up to the NLO-approximation. In principle,
their high order treatment is unjustifued. Nevertheless, one can assume that
these equations have the status of the kinetic equations of QCD studied
in this approach.

From one side, we understand that even if treated as kinetic equations these
equations are limited by our ignorance of the four-gluon interaction and
non-perturbative effects, by the simplified treatment of conservation laws etc.
Actually, the energy conservation is accounted by the $\ln x$
and $\ln (1-x)$ terms in the equations. In the perturbative expansion we 
cut off the Taylor expansions of the generating functions. Thus we approximate
the energy conservation. Namely this reveals itself in factorial moments
behaviour for small bins and in the oscillations of cumulant moments. 
In the computer solutions \cite{lo2, lo1, bfoc} the energy (but not $p_t$)
restrictions are
precisely considered and the results show better precision. Thus,
probably the inaccuracies of the analytic approach are connected just with
the improper treatment of the kinematic boundaries.

The modification of above equations was proposed \cite{eden} in
the framework of the dipole approach to QCD with more accurate kinematic bounds
accounting for the transverse momenta as well. It has been shown that the ratio
$r$ can be obtained in good agreement
with experimental data. Nevertheless, further study \cite{dede} of higher rank
moments of the multiplicity distribution predicted by the modified equations has
shown their extremely high sensitivity to higher orders of the perturbative
expansion. The results become inconclusive.

The more radical phenomenological approaches to generalize these equations
were attempted earlier \cite{123, 127, 128}.
In \cite{123} it was proposed to treat hadronization of partons at the
final stage of jet evolution in analogy with the ionization in
electromagnetic cascades where it results in their saturation and in the
finite length of the shower. This leads to some modified equations if the
analogy between ionization losses in QED and confinement in QCD is imposed.
Three different stages of the cascade were considered in the modified kinetic
equations proposed in \cite{127, 128}. No
quantitative results were, however, obtained.

Thus no successful generalization is at work nowadays. Rather, in view of
quite satisfactory agreement with experiment,  the general
theoretical trend has shifted to the direct
calculation of non-perturbative effects in some jet characteristics (see, e.g.,
\cite{yud, bdmz}) and to understanding effects described by the non-global
logarithms \cite{sala}.\\

5). {\bf The shortcomings of the analytic and numerical approaches}.
The success of numerical solutions of QCD equations \cite{21, lo2, lo1, bfoc}
raises the question if the generalization will give any other
noticable contribution. Our failure to describe more precisely the
ratio $r$ in analytic approach
could be just due some defects of the purely perturbative expansion at available
energies. The high order terms considered above correspond to corrections 
only due to more accurate treatment of the energy conservation and of the 
two-loop expression for the coupling strength (the term with $\beta _1$
in (\ref{al}), (\ref{mean}) considered). 
Also it was claimed recently \cite{hama} that the
renormalization group improvement of the perturbative results gives rise
to good description of experimental data.
Even in numerical calculations, it is still impossible to consider
in a proper way the transverse momenta. No high order terms have been added to
the kernels  (\ref{53}), (\ref{54}), (\ref{55}). The four-gluon vertex
has been completely ignored. Also, the non-perturbative effects are disregarded.
All these shortcomings provide the problems for
further studies in the framework of analytic and numerical approaches as well
as for Monte Carlo models. \\

In conclusion, I'd say that the practical accuracy of the pQCD calculations
is high enough. This is somewhat surprising in view of the rather large value
of the expansion parameter at present energies. They can serve as a good
estimate of the background in searches for new physics effects.
However some principal questions concerning the calculation of several
properties of quark-gluon jets and the validity of QCD equations for the
generating functions at higher orders has not yet been resolved.

This work has been supported in parts by the RFBR grants N 02-02-16779,
03-02-16134, NSH-1936.2003.2.

\end{document}